\documentclass{aastex}
\usepackage{spr-astr-addons}
\usepackage{url}

\RequirePackage{color}

\newcommand{\emaila}{jameel@giki.edu.pk}
\usepackage{latexsym}
\usepackage{graphics}
\usepackage{graphicx}
\usepackage{epsfig}

\begin{document}
\title{Unique first-forbidden $\beta$-decay rates for neutron-rich nickel isotopes in stellar environment}
\author{Jameel-Un Nabi \altaffilmark{1}}
\affil{Faculty of Engineering Sciences, GIK Institute of Engineering
Sciences and Technology, Topi 23640, Swabi, Khyber Pakhtunkhwa,
Pakistan} \email{\emaila} \and \author{Sabin Stoica} \affil{ Horia
Hulubei Foundation, P. O. Box MG-12, 071225, Magurele, Romania}
\altaffiltext{1}{Corresponding author email : jameel@giki.edu.pk}
\begin{abstract}
In astrophysical environments, allowed Gamow-Teller (GT) transitions
are important, particularly for $\beta$-decay rates in presupernova
evolution of massive stars, since they contribute to the fine-tuning
of the lepton-to-baryon content of the stellar matter prior to and
during the collapse of a heavy star. In environments where GT
transitions are unfavored, first-forbidden transitions become
important especially in medium heavy and heavy nuclei. Particularly
in case of neutron-rich nuclei, first-forbidden transitions are
favored primarily due to the phase-space amplification for these
transitions. In this work the total $\beta$-decay half-lives and the
unique first-forbidden(U1F) $\beta$-decay rates for a number of
neutron-rich nickel isotopes, $^{72-78}$Ni, are calculated using the
proton-neutron quasi-particle random phase approximation (pn-QRPA)
theory  in stellar environment for the first time. For the
calculation of the $\beta$-decay half-lives both allowed and unique
first-forbidden transitions were considered. Comparison of the total
half-lives is made with measurements and other theoretical
calculations where it was found that the pn-QRPA results are in
better agreement with experiments and at the same time are
suggestive of inclusion of rank 0 and rank 1 operators in
first-forbidden rates for still better results.
\end{abstract}

\section{Introduction}
Reliable and precise knowledge of the $\beta$-decay for neutron-rich
nuclei is crucial to an understanding of the $r$-process. Both the
element distribution on the r-path, and the resulting final
distribution of stable elements are highly sensitive to the
$\beta$-decay properties of the neutron-rich nuclei involved in the
process \cite{Kla83,Gro90}. There are about 6000 nuclei between the
$\beta$ stability line and the neutron drip line. Most of these
nuclei cannot be produced in terrestrial laboratories and one has to
rely on theoretical extrapolations for beta decay properties. In
neutron-rich environments electron neutrino captures could not only
amplify the effect of $\beta$-decays but the subsequent
$\nu$-induced neutron spallation  can also contribute towards
changing the r-abundance distribution pattern \cite{Mcl97}.
Correspondingly reliable predictions of $\beta$-decay for
neutron-rich nuclei are considered to be very important for
$r$-process nucleosynthesis.

The weak interaction rates are the important ingredients playing a
crucial role in practically all stellar processes: the hydrostatic
burning of massive stars, presupernova evolution of massive stars,
and nucleosynthesis (s-, p-, r-, rp-) processes (see, for example,
the seminal paper by Burbidge and collaborators \cite{Bur57}). For
densities  $\rho \lesssim 10^{11} g/cm^{3}$, stellar weak
interaction processes are dominated by Gamow-Teller (GT) and, if
applicable, by Fermi transitions. For nuclei lying in the vicinity
of the line of stability, forbidden transitions contribute sizably
for $\rho \gtrsim 10^{11} g/cm^{3}$ when the electron chemical
potential reaches values of the order of 30 MeV or more
\cite{Coo84}.

During presupernova stage of stellar evolution, electron capture
dominates the weak interaction processes. However, after silicon
depletion in the core and during the silicon shell burning,
$\beta$-decay competes temporarily with electron capture and further
cools the star. Unlike electron capture, the $\beta$-decay at
silicon burning stage increases and this bears consequences because
$\beta$-decays are additional neutrino source and add to the cooling
of the stellar core and a reduction in stellar entropy. This cooling
of stellar core can be quite efficient as often the average neutrino
energy in the involved $\beta$-decays is larger than for the
competing electron capture processes. Consequently, the
$\beta$-decay lowers the stellar core temperature after silicon
shell burning. For a discussion on the fine-tuning of
lepton-to-baryon ratio of stellar matter during presupernova
evolution see Ref. \cite{Auf94}.  As the density of the stellar core
increases the allowed $\beta$-decay becomes unimportant and hindered
due to the fact that the increasing Fermi energy of electrons blocks
the available phase space for the electron to be produced in the
$\beta$-decay. (For a detailed discussion of the role of weak
interaction in the presupernova evolution of massive stars see Ref.
\cite{Lan03}.) In case of neutron-rich nuclei, first-forbidden
$\beta$-decay may become important due to the enlarge phase space
for these transitions \cite{War88,War94}.

Reliable, quantitative estimates of $\beta$-decay half-lives of
neutron-rich nuclei are needed in astrophysics for the understanding
of supernova explosions, and the processes of nucleosynthesis,
particularly the $r$-process. The $\beta$-decay half-lives are also
needed for the experimental exploration of the nuclear landscape at
existing and future radioactive ion-beam facilities. The calculation
of $\beta$-decay half-lives in agreement with experimental results
has been a challenging problem for nuclear theorists.

Because of the scarcity of experimental data, majority of the
$\beta$-decay rates of the neutron-rich nuclei have been
investigated using theoretical models. Several models of different
level of sophistication for determining $\beta$-decay half-lives
have been proposed and applied over the years. One can mention the
more phenomenological treatments based on Gross Theory (e.g.
\cite{Tak73}) as well as microscopic treatments that employ the
proton-neutron quasiparticle random phase approximation (pn-QRPA)
\cite{Sta90,Hir93} or the shell model \cite{Mut91}. The later hybrid
version of the RPA models developed by M\"{o}ller and coworkers,
combines the pn-QRPA model with the statistical Gross Theory of the
first-forbidden decay (pnQRPA+ffGT)\cite{Moe03}. There are also some
models in which the ground state of the parent nucleus is described
by the Hartree-Fock BCS method, or other density functional method
(DF) and which use the continuum QRPA (CQRPA) (e.g. \cite{Eng99}).
Recently relativistic pn-QRPA (RQRPA) models have been applied in
the treatment of neutron-rich nuclei in the $N \sim$ 50, $N \sim$ 82
and $Z \sim$ 28 and 50 regions \cite{Nik05,Mar07}. Despite
continuing improvements the predictive power of these conventional
models is rather limited for $\beta$-decay half-lives of nuclei.

The microscopic calculations of allowed weak interaction rates using
the pn-QRPA model \cite{Kla84,Sta90,Hir93} and unique
first-forbidden (U1F) $\beta$-decay properties by Homma and
collaborators \cite{Hom96}, under terrestrial conditions, led to a
better understanding of the $r$-process. It was shown in Ref.
\cite{Hom96} that a larger contribution to the total transition
probability for near-stable and near-magic nuclei came from U1F
transitions (see Fig.~9 and Table IX of \cite{Hom96}). According to
the $\beta$-decay studies by Borzov \cite{Bor06}, the allowed
$\beta$-decay approximation alone is not adequate to describe the
isotopic dependence of the $\beta$-decay characteristics, specially
for the nuclei crossing the closed $N$ and $Z$ shells  and forbidden
transitions give a dominant contribution to the total half-life for
nuclei crossing the closed shells, specially for $N  < $ 50 in
$^{78}$Ni region. Recently large-scale shell-model calculation of
the half-lives, including first-forbidden contributions ,for
$r$-process waiting-point nuclei was also performed \cite{Zhi13}.
However there was a need to go to domains of high temperature and
density scales where the weak interaction rates are of decisive
importance in studies of the stellar evolution (see e.g.
\cite{Bor06}). Moving on from terrestrial to stellar domain one
notes two microscopic approaches, namely the shell model
\cite{Lan00} and the pn-QRPA \cite{Nab04}, have so far been
successfully used in the large-scale calculation of stellar
weak-interaction mediated rates for $r$-process applications.
Whereas the advantage of the shell model clearly lies in it's
ability to take into account the detailed structure of the
$\beta$-strength functions, the shell model is forced to use
approximations like Brink's hypothesis (in electron capture
direction) and back resonances (in $\beta$-decay direction) when
extending the calculation to high-lying parent excited states which
can contribute rather significantly to the total weak rate in
stellar environment. Brink's hypothesis states that GT strength
distribution on excited states is \textit{identical} to that from
ground state, shifted \textit{only} by the excitation energy of the
state. GT back resonances are the states reached by the strong GT
transitions in the inverse process (electron capture) built on
ground and excited states. The pn-QRPA model, on the other hand,
gets rid of such approximations and allows a state-by-state
evaluation of the weak rates by summing over Boltzmann-weighted,
microscopically determined GT strengths \textit{for all} parent
excited states.

The recent measurement of the GT$_{+}$ strength distribution of
$^{76}$Se \cite{Gre08} supports the argument that, due to nuclear
correlations across the $N = 40$ shell gap, the GT transitions for
fp-nuclei with proton numbers $Z < 40$ and neutron numbers $N > 40$
would not be Pauli blocked. Nevertheless for neutron-rich nuclei U1F
transitions are favored due to the amplification of available phase
space. Further in the large-scale shell-model first-forbidden
calculation of Zhi and collaborators \cite{Zhi13} the authors note
that for $N$ = 50 nuclei with $Z \ge$ 28, contributions to the first
forbidden transitions are nearly exclusively due to rank 2
operators. In this paper we attempt to calculate the allowed as well
as the unique first-forbidden (U1F) $\beta$-decay rates for
neutron-rich nickel isotopes ($^{72-78}$Ni) using the pn-QRPA model.
Further the calculations are extended from terrestrial to stellar
domain for the first time using the pn-QRPA theory. Motivation of
the current work was primarily based on the comments given by Homma
and collaborators that the U1F transition has large enough
contribution to the total transition probability and inclusion of
U1F transition greatly improves the calculated half-lives using the
pn-QRPA model \cite{Hom96}.  It is worthwhile to also study the
effects of non-unique forbidden transitions which are expected, for
some nuclei, to have still larger contribution to the calculated
half-lives (as also mentioned in \cite{Hom96}). We would like to
include this desired modification in our nuclear code as a future
project. The theory of allowed and first-forbidden $\beta$-decay
transitions (along with its leptonic phase-space content and nuclear
matrix elements) is well-established
\cite{Sch66,Wu66,Com73,Bla79,Beh82}. Whereas the allowed
$\beta$-decay (having the same parity before and after the decay) is
relatively simple to calculate, the first-forbidden decay shows a
far wider spectrum both in lepton kinematics and in nuclear matrix
elements \cite{Wu66,Sch66,Beh82}.

The pn-QRPA formalism for calculations of the total $\beta$-decay
half-lives and U1F $\beta$-decay rates for neutron-rich nickel
isotopes are discussed in Section 2. Results and comparisons with
experimental and other theoretical calculation are shown in Section
3. Section 4 outlines the importance of the results in connection
with the simulation of the advanced stages of stellar evolution and
heavy element nucleosynthesis and concludes our discussion on
consequences of U1F $\beta$-decays in stellar matter.

\section{Formalism}

In the pn-QRPA formalism \cite{Mut92}, proton-neutron residual
interactions occur in two different forms, namely as particle-hole
($ph$) and particle-particle ($pp$) interactions. The particle-hole
and particle-particle interactions can be given a separable form
because $\beta$$^{-}$ transitions are dominated by the particle-hole
interaction. These $ph$ and $pp$ forces are repulsive and
attractive, respectively, when the strength parameters $\chi$ for
$ph$ GT force and $\kappa$ for $pp$ GT force take positive values.
The advantage of using these separable GT forces is that the QRPA
matrix equation reduces to an algebraic equation of fourth order,
which is much easier to solve as compare to full diagonalization of
the non-Hermitian matrix of large dimensionality \cite{Mut92,Hom96}.

Essentially we first construct a quasiparticle basis (defined by a
Bogoliubov transformation) with a pairing interaction, and then
solve the RPA equation with a schematic separable GT residual
interaction. The single particle energies were calculated using a
deformed Nilsson oscillator potential with a quadratic deformation.
The pairing correlation was taken into account in the BCS
approximation using constant pairing forces. The BCS calculation was
performed in the deformed Nilsson basis for neutrons and protons
separately. The formalism for calculation of allowed $\beta$-decay
rates in stellar matter using the pn-QRPA model can be seen in
detail from Refs. \cite{Nab99,Nab04}. Below we describe briefly the
necessary formalism to calculate the U1F $\beta$-decay rates.

Allowed Fermi and GT transitions have a maximum spin change of one
unit  and no change in parity of the wavefunction. In contrast, the
most frequent occurrence of a forbidden decay is when the initial
and final states have opposite parities, and thus the selection rule
for allowed decay is violated. The spins of the initial and final
states can be different at most by $\Delta$J= 0, $\pm$1, $\pm$2 for
the first-forbidden transitions (those transitions for which $\Delta
J^{\pi} = 2^{-}$ are termed as unique first-forbidden). The isospin
selection rule remains the same as for allowed decays. For the
calculation of the U1F $\beta$-decay rates, nuclear matrix elements
of the separable forces which appear in RPA equation are given by

\begin{equation}
V^{ph}_{pn,p^{\prime}n^{\prime}} = +2\chi
f_{pn}(\mu)f_{p^{\prime}n^{\prime}}(\mu),
\end{equation}

\begin{equation}
V^{pp}_{pn,p^{\prime}n^{\prime}} = -2\kappa
f_{pn}(\mu)f_{p^{\prime}n^{\prime}}(\mu),
\end{equation}

where
\begin{equation}
f_{pn}(\mu)=<j_{p}m_{p}|t_{-}r[\sigma Y_{1}]_{2\mu}|j_{n}m_{n}>
\end{equation}
is a single-particle U1F transition amplitude (the symbols have
their normal meaning). Note that $\mu$ takes the values
$\mu=0,\pm1$, and $\pm2$ (for allowed decay rates $\mu$ only takes
the values $0$ and $\pm1$), and the proton and neutron states have
opposite parities \cite{Hom96}. For the sake of consistency we chose
to keep the same values of interaction constants ($\chi$ = 0.35 MeV
and $\kappa$ = 0.001 MeV  for allowed and $\chi$ = 0.35 MeV
fm$^{-2}$ and $\kappa$ = 0.001 MeV fm$^{-2}$ for U1F transitions).
For choice of strength of interaction constants in the pn-QRPA model
we refer to \cite{Sta90,Hir93,Hom96}.

Deformations of the nuclei were calculated using
\begin{equation}
\delta = \frac{125(Q_{2})}{1.44 (Z) (A)^{2/3}},
\end{equation}
where $Z$ and $A$ are the atomic and mass numbers, respectively and
$Q_{2}$ is the electric quadrupole moment taken from Ref.
\cite{Moe81}. Q-values were taken from the recent mass compilation
of Audi et al. \cite{Aud03}.

The U1F $\beta$-decay rates from the $\mathit{i}$th state of the
parent to the $\mathit{j}$th state of the daughter nucleus is given
by

\begin{equation}
\lambda_{ij} = \frac{m_{e}^{5}c^{4}}{2\pi^{3}\hbar^{7}}\sum_{\Delta
J^{\pi}}g^{2}f(\Delta J^{\pi};ij)B(\Delta J^{\pi};ij),
\end{equation}
where $f(\Delta J^{\pi};ij)$ and $ B(\Delta J^{\pi};ij)$ are the
integrated Fermi function and the reduced transition probability,
respectively, for the transition $i \rightarrow j$ which induces a
spin-parity change $\Delta J^{\pi}$ and $g$ is the weak coupling
constant which takes the value $g_{V}$ or $g_{A}$ according to
whether the $\Delta J^{\pi}$ transition is associated with the
vector or axial-vector weak-interaction. The phase-space factors
$f(\Delta J^{\pi};ij)$ are given as integrals over the lepton
distribution functions and hence are sensitive functions of the
temperature and density in stellar interior. The $B(\Delta
J^{\pi};ij)$ are related to the U1F weak interaction matrix elements
stated earlier.

For the first-forbidden transitions the integral can be obtained as

\begin{eqnarray}
f = \int_{1}^{w_{m}} w \sqrt{w^{2}-1}
(w_{m}-w)^{2}[(w_{m}-w)^{2}F_{1}(Z,w) \nonumber\\
+ (w^{2}-1)F_{2}(Z,w)] (1-G_{-}) dw,
\end{eqnarray}
where $w$ is the total kinetic energy of the electron including its
rest mass and $w_{m}$ is the total $\beta$-decay energy ($ w_{m} =
m_{p}-m_{d}+E_{i}-E_{j}$, where $m_{p}$ and $E_{i}$ are mass and
excitation energies of the parent nucleus, and $m_{d}$ and $E_{j}$
of the daughter nucleus, respectively). $G_{-}$ are the electron
distribution functions. Assuming that the electrons are not in a
bound state, these are the Fermi-Dirac distribution functions,
\begin{equation} G_{-} = [exp
(\frac{E-E_{f}}{kT})+1]^{-1}.
\end{equation}
Here $E=(w-1)$ is the kinetic energy of the electrons, $E_{f}$ is
the Fermi energy of the electrons, $T$ is the temperature, and $k$
is the Boltzmann constant.

The Fermi functions, $F_{1}(\pm Z,w)$ and $F_{2}(\pm Z,w)$ appearing
in Eq.~(6) are calculated according to the procedure adopted by Gove
and Martin \cite{Gov71}.

The number density of electrons associated with protons and nuclei
is $\rho Y_{e} N_{A}$, where $\rho$ is the baryon density, $Y_{e}$
is the ratio of electron number to the baryon number, and $N_{A}$ is
the Avogadro's number.
\begin{equation}
\rho Y_{e} =
\frac{1}{\pi^{2}N_{A}}(\frac {m_{e}c}{\hbar})^{3} \int_{0}^{\infty}
(G_{-}-G_{+}) p^{2}dp,
\end{equation}
where $p=(w^{2}-1)^{1/2}$ is
the electron or positron momentum, and Eq.~(8) has the units of
\textit{moles $cm^{-3}$}. $G_{+}$ are the positron distribution
functions given by
\begin{equation} G_{+} =\left[\exp
\left(\frac{E+2+E_{f} }{kT}\right)+1\right]^{-1}.
\end{equation}
Eq.~(8) is used for an iterative calculation of Fermi energies for
selected values of $\rho Y_{e}$ and $T$.

There is a finite probability of occupation of parent excited states
in the stellar environment as a result of the high temperature in
the interior of massive stars. Weak decay rates then also have a
finite contribution from these excited states. The occupation
probability of a state $i$ is calculated on the assumption of
thermal equilibrium,

\begin{equation} P_{i} = \frac {exp(-E_{i}/kT)}{\sum_{i=1}exp(-E_{i}/kT)}, \end{equation}
where $E_{i}$ is the excitation energy of the state $i$,
respectively. The rate per unit time per nucleus for any weak
process is then given by
\begin{equation} \lambda = \sum_{ij}P_{i}
\lambda_{ij}.
\end{equation} The summation over all initial and final states are carried out
until satisfactory convergence in the rate calculations is achieved.
We note that due to the availability of a huge model space (up to 7
major oscillator shells) convergence is easily achieved in our rate
calculations for excitation energies well in excess of 10 MeV (for
both parent and daughter states).

\section{Results and comparison}

In this section we present the phase space calculation, strength
distributions calculation, total $\beta$-decay half-lives and total
stellar $\beta$-decay rates for neutron-rich Ni isotopes.  The
calculated total terrestrial half-lives are also compared with the
measured ones to estimate the accuracy of the model.

The phase space calculation for allowed and U1F transitions, as a
function of stellar temperature and density, for the neutron-rich
nickel isotopes are shown in Tables~1 and 2. The phase space is
calculated at selected density of 10$^{2}$ g/cm$^{3}$, 10$^{6}$
g/cm$^{3}$ and 10$^{10}$ g/cm$^{3}$ (corresponding to low,
intermediate and high stellar densities, respectively) and stellar
temperature T$_{9}$ = 0.01, 1, 3 and 10 (T$_{9}$ gives the stellar
temperature in units of 10$^{9}$ K). Table~1 shows the result for
$^{72-75}$Ni whereas Table~2 displays the phase space calculation
for $^{76-78}$Ni. It can be seen from these tables that for low and
intermediate stellar densities the phase space increase by 2-5
orders of magnitude as the stellar temperature goes from T$_{9}$ =
0.01 to 1. Moreover the phase space remains same as stellar
temperature rises from T$_{9}$ = 1 to 10. For high densities the
phase space is essentially zero at T$_{9}$ = 0.01 and increases with
increasing temperatures. As the stellar core becomes more and more
dense the phase space decreases. For the case of $^{72-75}$Ni, the
phase space increases by 1-2 orders of magnitude for U1F transitions
as the stellar temperature and density increases.  As the nickel
isotopes becomes more and more neutron-rich, the phase space
enhancement for U1F transitions decreases but still is bigger than
the phase space for allowed transitions.

In order to ensure reliability of calculated  $\beta$-decay rates,
experimental (XUNDL) data were incorporated in the calculation
wherever possible and the calculated excitation energies were
replaced with measured levels when they were within 0.5 MeV of each
other. Missing measured states were inserted and inverse transitions
(along with their log$ft$ values), where applicable, were also taken
into account. No theoretical levels were replaced with the
experimental ones beyond the excitation energy for which
experimental compilations had no definite spin and/or parity. This
recipe for incorporation of measured data is same as used in earlier
pn-QRPA calculation of stellar weak rates \cite{Nab99,Nab04}.

\begin{center}
\begin{figure}[!htb]
  \begin{center}
  \includegraphics[width=0.8\textwidth]{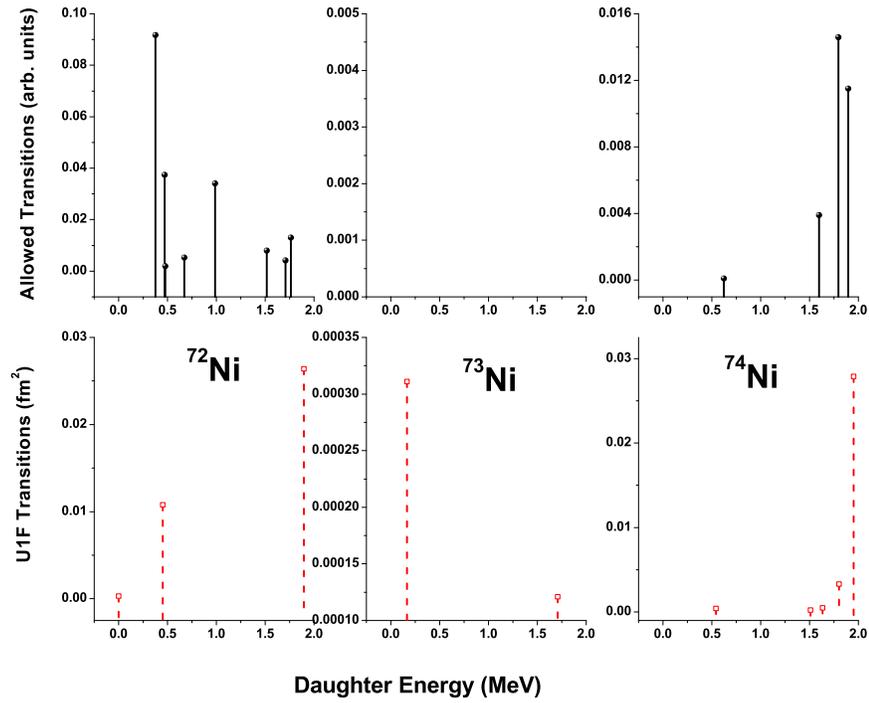}
  \end{center}
\caption{\small Calculated strength distributions for allowed and
U1F transitions for $^{72-74}$Ni.} \label{fig0a}
\end{figure}
\end{center}

\begin{center}
\begin{figure}[!htb]
  \begin{center}
  \includegraphics[width=0.8\textwidth]{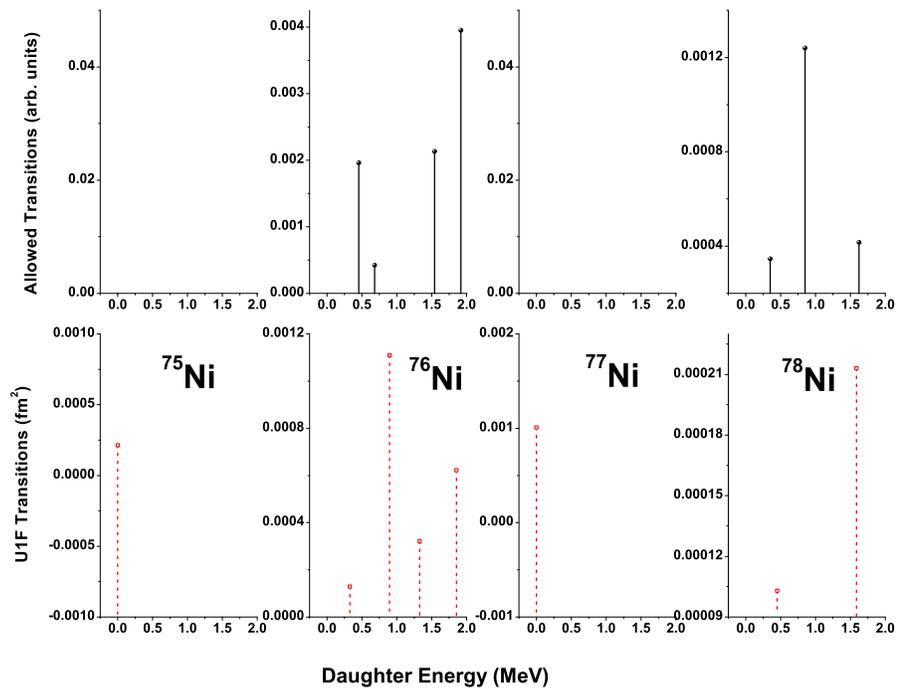}
  \end{center}
\caption{\small Same as Fig.~\ref{fig0a} but for $^{75-78}$Ni.}
\label{fig0b}
\end{figure}
\end{center}

We present the calculated strength distributions of $^{72-78}$Ni,
both due to allowed and U1F transitions in Fig.~\ref{fig0a} and
Fig.~\ref{fig0b}. Calculations for $^{72-74}$Ni are displayed in
Fig.~\ref{fig0a} whereas Fig.~\ref{fig0b} depicts result for
$^{75-78}$Ni. All energies are given in units of MeV, allowed
strengths are given in arbitrary units and U1F transitions are given
in units of fm$^{2}$. In these figures we only show low-lying
strength distribution up to an excitation energy of 2 MeV in
daughter nucleus. Calculated values of strength smaller than
10$^{-4}$ are not shown in these figures. For the case of $^{72}$Ni,
we show the 1$^{+}$ states as filled circles and 2$^{-}$ states as
open squares
 (Fig.~\ref{fig0a}). For the case of $^{73}$Ni, no allowed GT
transitions were calculated up to excitation energies of 2 MeV in
$^{73}$Cu and low-energy region is populated by U1F transitions.
Same is the case for  other odd-A isotopes of nickel, $^{75}$Ni and
$^{77}$Ni (Fig.~\ref{fig0b}).  The excitation energy of 1$^{+}$
state is important for the evaluation of the transition rate in
$^{78}$Ni and bear consequences for $r$-process simulation as
pointed out in a recent work by Minato and Bai \cite{Min13}. The
effect of tensor force on $\beta$-decay rates was studied within the
Hartree-Fock plus proton-neutron random phase approximation
employing the Skyrme force in Ref. \cite{Min13}. The authors
concluded that the tensor force makes dramatic improvement in the
prediction of $\beta$-decay half-lives of the even-even magic (and
semi-magic) nuclei where the effect of isoscalar pairing is
negligible. It was further deduced that the tensor force had a
significant contribution to the low-lying GT distribution, in
particular, the effect of the tensor terms in the residual
interaction of RPA was more important than the change of the
spin-isospin splitting in the Hartree-Fock. Both Figs.~~\ref{fig0a}
and ~\ref{fig0b} demonstrate the fact that U1F transitions play a
crucial role in reducing the energy of the GT state (specially for
the odd A nuclei) which in turn contribute to the reduction of
calculated half-lives as discussed below.

\begin{center}
\begin{figure}[!htb]
  \begin{center}
  \includegraphics[width=0.8\textwidth]{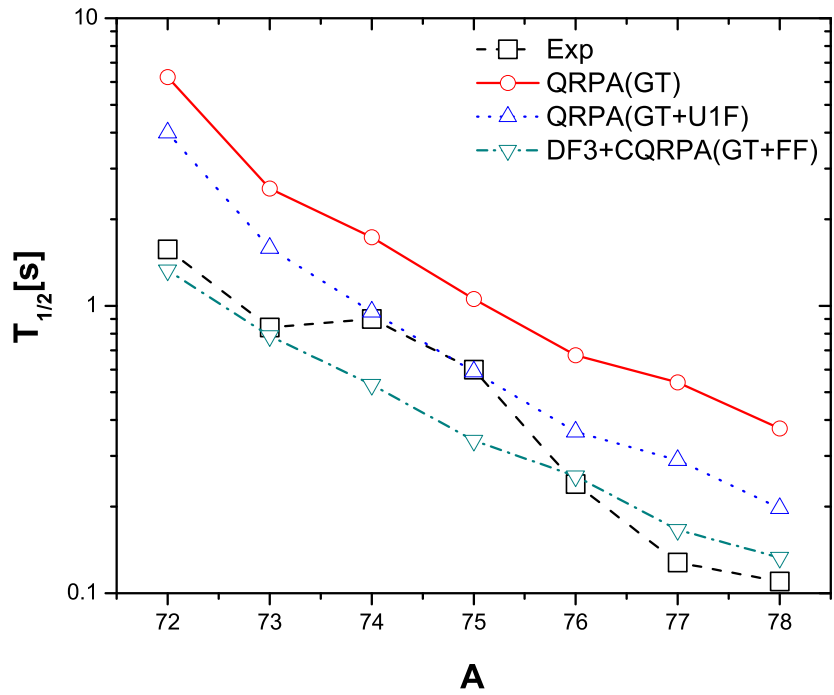}
  \end{center}
\caption{\small Total $\beta$-decay half-lives for Ni isotopes
calculated from the pn-QRPA (this work) including only the allowed
(GT), allowed  plus unique first-forbidden (GT+U1F) transitions, in
comparison with the DF3+CQRPA \cite{Bor05} and experimental data
\cite{Fra98,Ame98,Hos05}.} \label{fig1}
\end{figure}
\end{center}

For the Ni isotopic chain the results for terrestrial half-lives are
shown in Fig.~\ref{fig1}. We also compare our calculation with the
experimental values \cite{Fra98,Ame98,Hos05} and with the
self-consistent density functional + continuum quasiparticle random
phase approximation (DF3 + CQRPA) calculation of Ref. \cite{Bor05}.
The pn-QRPA calculated half-lives including only allowed GT
transitions are also shown in Fig.~\ref{fig1} to show the
improvement brought by inclusion of U1F $\beta$-decay rates which
results in calculation of smaller half-lives. Excellent agreement
with measured half-lives is achieved when U1F transitions are
included for the case of $^{74,75}$Ni. It is to be noted that Ref.
\cite{Bor05} overestimate the total $\beta$-decay rates for
$^{74,75}$Ni roughly by a factor of two. The pn-QRPA half-lives are
roughly a factor two bigger than experimental half-lives
\cite{Fra98} for $^{72,73}$Ni whereas the calculated half-lives of
Ref. \cite{Bor05} are in better comparison with the corresponding
measured values. For the remaining cases, $^{76-78}$Ni, the pn-QRPA
calculated half-lives are again roughly a factor two bigger than the
measured half-lives. The DF3 + CQRPA calculation of Ref.
\cite{Bor05} are also 20$\%$ - 30$\%$ bigger than the measured
half-lives for the case of $^{77-78}$Ni. The results are suggestive
that possible inclusion of rank 0 and rank 1 operators in
calculation of first-forbidden $\beta$-decay rates can bring further
improvement in the overall comparison with the experimental data
which we plan to incorporate in future. The incorporation of U1F
transitions decreased the calculated half-lives from about 36$\%$
for the case of $^{72}$Ni to about 47$\%$ for the case of $^{78}$Ni.
Fig.~\ref{fig1} proves that the inclusion of the U1F contribution
makes the half-life predictions more reliable for neutron-rich
nickel isotopes.

Recently a large-scale shell-model calculation of half-lives
including first-forbidden transitions (including rank 0, 1 and 2
operators) was performed for $r$-process waiting-point nuclei
\cite{Zhi13}. In their work the authors remark that for $N$ = 50
nuclei with $Z \ge$ 28, contributions to the first forbidden
transitions are nearly exclusively due to rank 2 operators. There
the authors concluded that forbidden transitions have a small effect
(reduction of 5$\%$ - 25$\%$)  on the half-lives of $N$ = 50
waiting-point nuclei which is much smaller than the contributions
from forbidden transitions calculated in this work. However there
are at least three big differences in the shell model and the
current pn-QRPA model calculations. Firstly, as mentioned
previously, the current pn-QRPA model is not capable of calculating
forbidden contributions of rank 0 and rank 1 operators which the
shell model calculation can (it is remarked by the authors that for
$Z <$ 28  and $N$ = 50, rank 0 and rank 1 operators contribute
significantly to the first forbidden transitions). Secondly
quenching was used in the shell model calculation and different
quenching factors ranging from 0.38 to 1.266 was used for the
various nuclear matrix elements. No explicit quenching factor was
used in current calculation in contrast to shell model calculation.
Last but not the least, due to model space limitation in shell model
calculation, authors in Ref. \cite{Zhi13} were not able to recover
the full GT and first-forbidden strengths built on the ground and
isomeric states. A completely converged calculation of the
first-forbidden transition strength was not possible in shell model
due to computational limitations (they used a Lanczos scheme with
100 iterations which was able to converge the states for excitation
energies in the vicinity of only 2.5 MeV). The pn-QRPA model has no
such computational limitations and convergence was achieved for
excitation energies well in excess of 10 MeV. The U1F contribution
in the current pn-QRPA model is well within the percentage
contribution of first-forbidden transitions (to the total
half-lives) as calculated by authors employing the FRDM + QRPA model
in Ref. \cite{Moe03}.

\begin{center}
\begin{figure}[!htb]
  \begin{center}
  \includegraphics[width=0.8\textwidth]{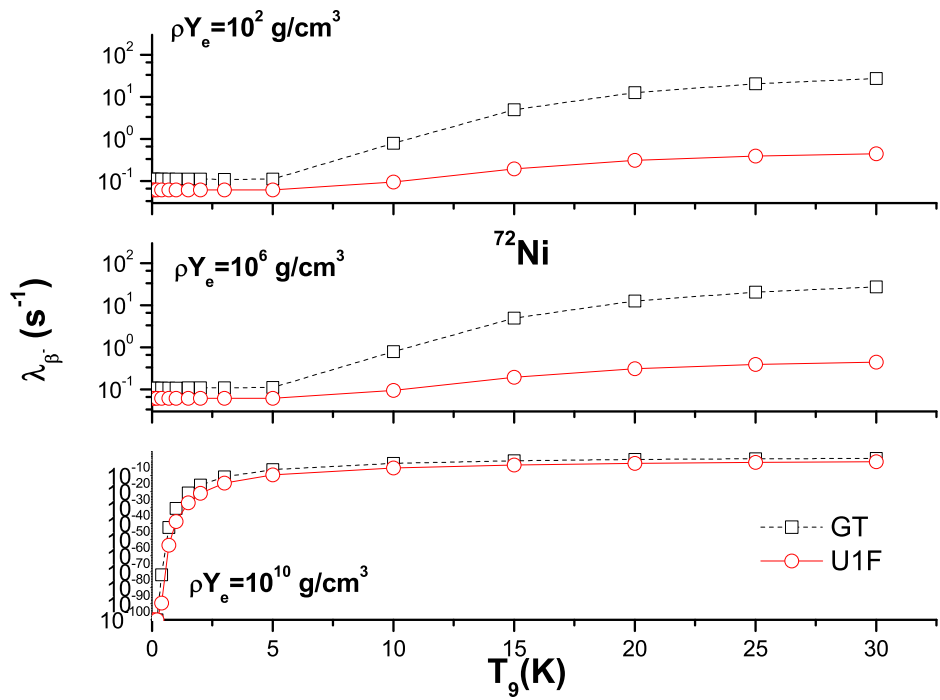}
  \end{center}
\caption{\small Allowed (GT) and  unique first-forbidden (U1F)
$\beta$-decay rates on $^{72}$Ni as function of temperature for
different selected densities. All $\beta$ decay rates are given in
units of sec$^{-1}$. Temperatures (T$_{9}$) are given in units of
10$^{9}$ K. } \label{fig2}
\end{figure}
\end{center}

\begin{center}
\begin{figure}[!htb]
  \begin{center}
  \includegraphics[width=0.8\textwidth]{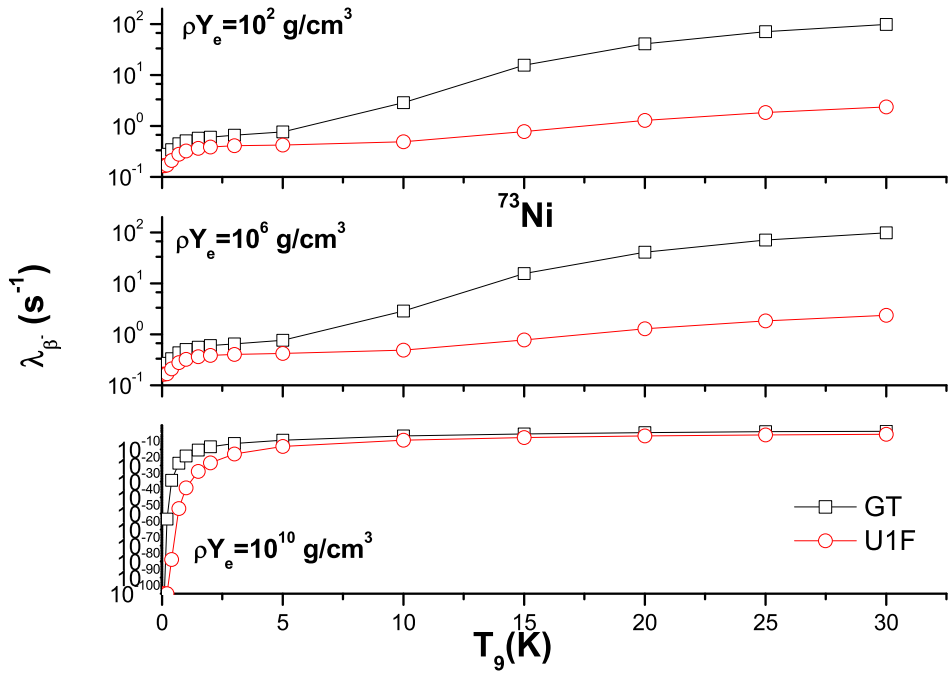}
  \end{center}
\caption{\small Same as Fig.~\ref{fig2} but for $^{73}$Ni.}
\label{fig3}
\end{figure}
\end{center}

\begin{center}
\begin{figure}[!htb]
  \begin{center}
  \includegraphics[width=0.8\textwidth]{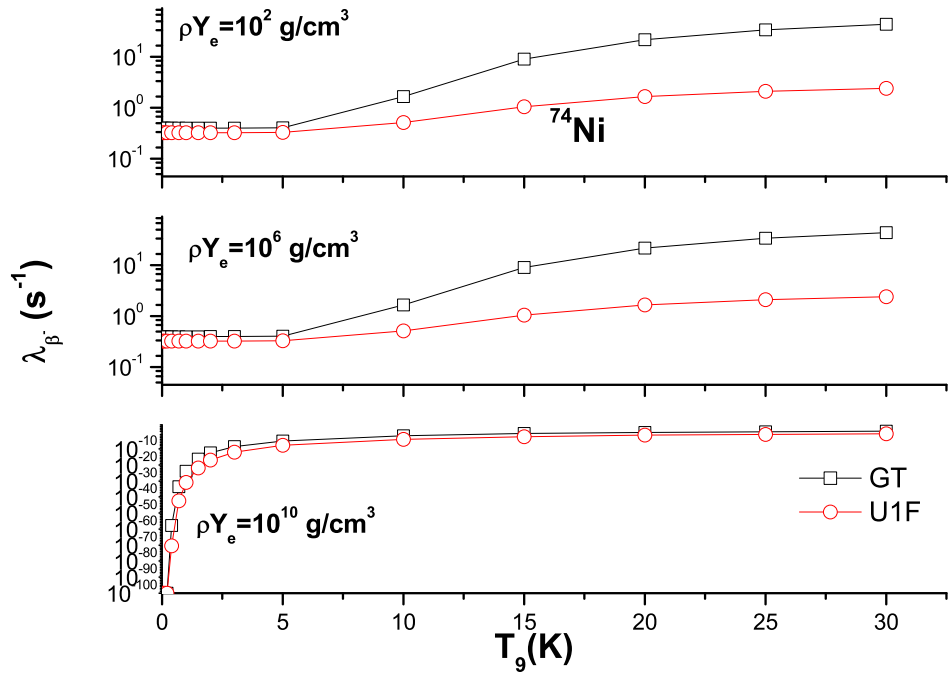}
  \end{center}
\caption{\small Same as Fig.~\ref{fig2} but for $^{74}$Ni.}
\label{fig4}
\end{figure}
\end{center}

\begin{center}
\begin{figure}[!htb]
  \begin{center}
  \includegraphics[width=0.8\textwidth]{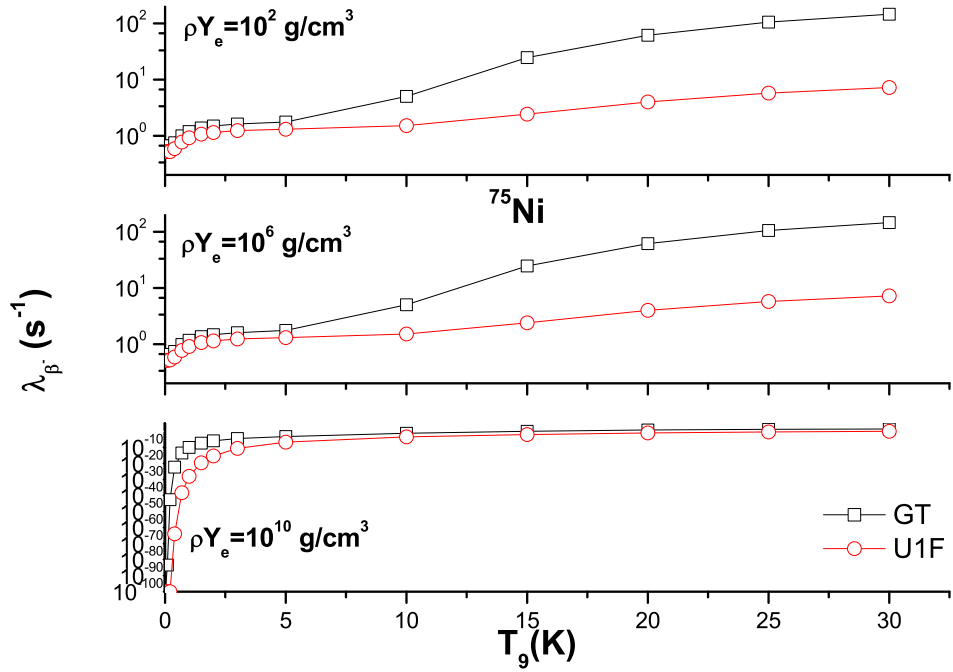}
  \end{center}
\caption{\small Same as Fig.~\ref{fig2} but for $^{75}$Ni.}
\label{fig5}
\end{figure}
\end{center}

\begin{center}
\begin{figure}[!htb]
  \begin{center}
  \includegraphics[width=0.8\textwidth]{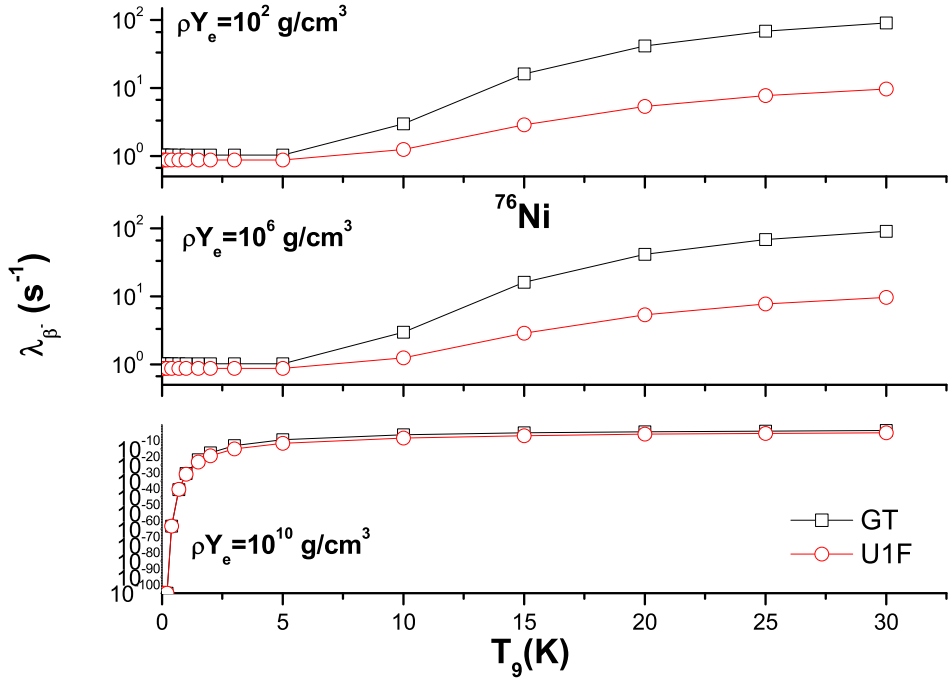}
  \end{center}
\caption{\small Same as Fig.~\ref{fig2} but for $^{76}$Ni.}
\label{fig6}
\end{figure}
\end{center}

\begin{center}
\begin{figure}[!htb]
  \begin{center}
  \includegraphics[width=0.8\textwidth]{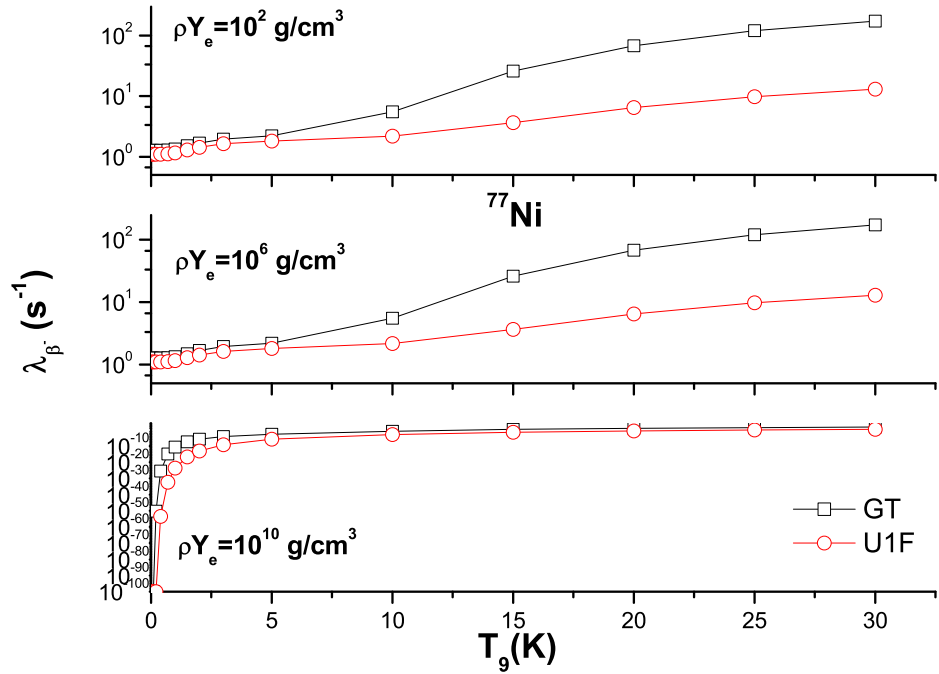}
  \end{center}
\caption{\small Same as Fig.~\ref{fig2} but for $^{77}$Ni.}
\label{fig7}
\end{figure}
\end{center}

\begin{center}
\begin{figure}[!htb]
  \begin{center}
  \includegraphics[width=0.8\textwidth]{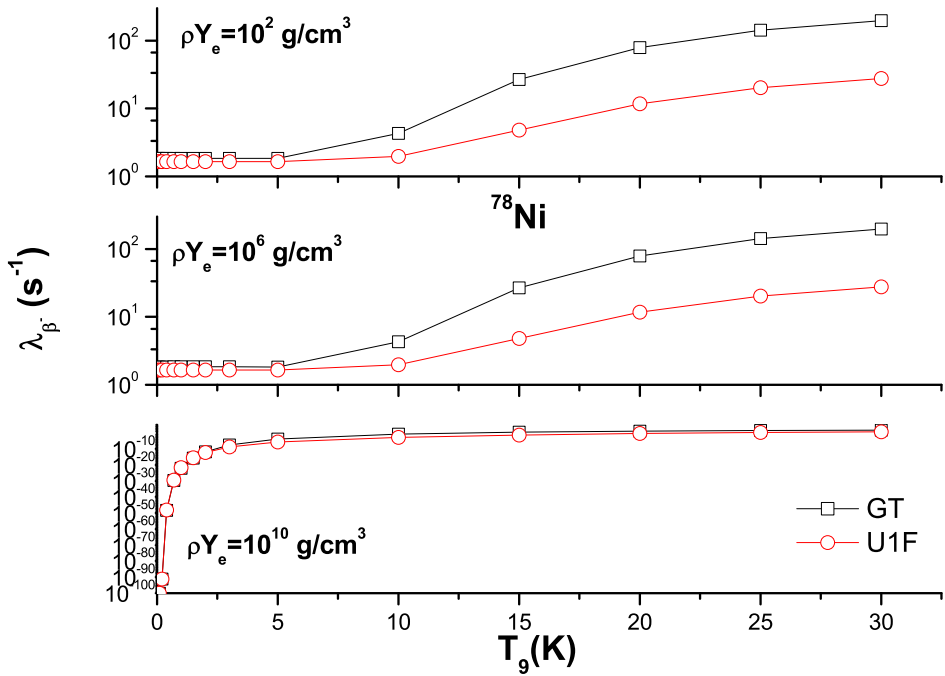}
  \end{center}
\caption{\small Same as Fig.~\ref{fig2} but for $^{78}$Ni.}
\label{fig8}
\end{figure}
\end{center}

The U1F $\beta$-decay rates were calculated for densities in the
range 10-10$^{11}$g/cm$^{3}$ and temperature range 0.01 $\leq$
T$_{9}$ $\leq$ 30. Figs.~\ref{fig2}-~\ref{fig8} show three panels
depicting pn-QRPA calculated allowed and U1F $\beta$-decay rates for
temperature range 0.01 $\leq$ T$_{9}$ $\leq$ 30 for the neutron-rich
nickel isotopes.  The upper panel depicts the situation at a low
stellar density of 10$^{2}$ g/cm$^{3}$, the middle panel for
10$^{6}$ g/cm$^{3}$ and the lower panel at a high stellar density of
10$^{10}$ g/cm$^{3}$.  It is to be noted that for all figures the
abscissa is given in logarithmic scales. Fig.~\ref{fig2} depicts
stellar $\beta$-decay rates for $^{72}$Ni in units of $s^{-1}$. It
is pertinent to mention that contribution from all excited states
are included in the final calculation of all decay rates. It can be
seen from this figure that for low and intermediate densities
(10$^{2}$ and 10$^{6}$ g/cm$^{3}$) the $\beta$-decay rates (both
allowed and U1F) remain almost the same as a function of stellar
temperature. This feature remains the same for all following
figures. At low temperatures (T$_{9}$ $\leq$ 5) the U1F transition
is roughly half the allowed transition (for phase space
consideration refer to Table~1). At high stellar temperatures the
allowed transitions become bigger by as much as a factor of 60 as
more and more excited states contribute significantly to the allowed
GT rates.  At high stellar density (10$^{10}$ g/cm$^{3}$) the
$\beta$-decay rates tend to go down due to the blocking of the
available phase space of increasingly degenerate electrons and one
notes that allowed GT rates are tens of orders of magnitude bigger
than UIF rates. As the stellar temperature rises, the orders of
magnitude difference reduces and for T$_{9}$ = 30 the allowed rates
are around two orders of magnitude bigger than U1F rates. At high
densities and high temperatures the contribution to the total
$\beta$-decay rates by the excited states is very important. The
rates rise with higher temperature as more excited states start
contributing when the temperature increases. Increasing temperature
and density have opposite contributions to the calculated rates.
Whereas growth of the stellar density suppresses the $\beta$ decay
rates due to a diminishing phase space available for escaping
electrons, increase in temperature weakens the Pauli blocking and
therefore enhances the contribution of the GT$_{-}$ transitions from
excited states of the parent nucleus.

Fig.~\ref{fig3} shows the case of odd A nucleus $^{73}$Ni. For low
and intermediate stellar densities the allowed GT rates are bigger
by a factor of 1.6 at T$_{9}$ = 0.01 to a factor of 42 at T$_{9}$ =
30. For high densities the allowed rates are a factor 90 bigger than
the U1F rates at T$_{9}$ = 30. Due to a significantly larger number
of available parent excited states in odd-A nuclei, the
contributions from allowed and U1F rates increases with temperature
when compared with the even-even isotopes. The results almost follow
the same trend as shown in Fig.~\ref{fig2}. The effects of
increasing temperature and density on the decay rates were discussed
earlier.

Fig.~\ref{fig4} shows the next even-even case of  $^{74}$Ni. Here
one notes that at low and intermediate stellar densities and low
temperatures (T$_{9}$ $\leq$ 5) the allowed GT rates are only a
factor of 1.2 bigger than UIF rates whereas at T$_{9}$ = 30 the
allowed rates are a factor 18 bigger. For high densities the allowed
rates are a factor 38 bigger than the U1F rates at a stellar
temperature of T$_{9}$ = 30. Once again the results almost follow
the same trend as shown in Fig.~\ref{fig2}.

Similarly the allowed GT rates for $^{75}$Ni are only a factor of
1.3 bigger than U1F rates for low temperatures (T$_{9}$ $\sim$ 0.2)
and low and intermediate stellar densities (Fig.~\ref{fig5}). As the
stellar temperatures increases to T$_{9}$ = 30, the allowed rates
are bigger by a factor of 20. For high densities and temperature the
allowed rates are almost a factor 40 bigger.

Fig.~\ref{fig6} shows the  case of  $^{76}$Ni and is very much
similar to Fig.~\ref{fig4}. At T$_{9}$ = 30 the enhancement in total
$\beta$-decay rates, due to U1F contributions, is roughly double
that for the case of $^{74}$Ni (Fig.~\ref{fig4}). Similarly
Fig.~\ref{fig7} for $^{77}$Ni is akin to Fig.~\ref{fig5}. One notes
that, for both even-even and odd-A isotopes of nickel, the U1F
contributions increases as the number of neutrons increases. This is
primarily because of the reduction in phase space of allowed
$\beta$-decay rates as $N$ increases.

Fig.~\ref{fig8} for $^{78}$Ni (double-magic nucleus) is very much
similar to the even-even case of $^{76}$Ni (Fig.~\ref{fig6}). The
half-life of $^{78}$Ni (and consequently it's $\beta$-decay rate) is
believed to set the processing timescale for the synthesis of
elements beyond A $\sim$ 80. The neutron separation energy shows
discontinuity at $N$ = 50 and consequently the $r$-process matter
flow slows down. The $\beta$-decay of $^{78}$Ni then takes charge
and accordingly $^{78}$Ni is considered to be a crucial waiting
point nuclei for $r$-process. Fig.~\ref{fig8} shows that for stellar
density range 10$^{2}$- 10$^{6}$ g/cm$^{3}$, the total $\beta$-decay
rates of $^{78}$Ni increase from 3.5 $s^{-1}$ at T$_{9}$ = 0.01 to
around 225 $s^{-1}$ at T$_{9}$ = 30. As the stellar core stiffens to
a density of around 10$^{10}$ g/cm$^{3}$, the $\beta$-decay rates
decrease by tens of orders of magnitude at low stellar temperatures
to around 67 $s^{-1}$ at T$_{9}$ = 30. Further our calculation shows
significant U1F contribution to total $\beta$-decay rates of
$^{78}$Ni at low stellar temperatures T$_{9} \le$ 1. At high
temperatures of T$_{9}$ = 30, the allowed rates are a factor 7 (for
low and intermediate densities) to factor 13 bigger (for high
densities) than the U1F $\beta$-decay rates.

It is to be noted that increasing the stellar temperature weakly
affects the U1F $\beta$-decay rates. The reason is that the
first-forbidden transitions are of relatively high energies and
basically of the particle-hole nature. Therefore a smearing of
single-particle strengths due to pairing correlations and
temperature influence them considerably less. On the other hand the
allowed GT rates increase at a faster pace with increasing
temperature and hence explains this feature in all figures.

In the $\beta$-decay of a particular isotope, an emitted electron
can receive any kinetic energy from zero up to the maximum kinetic
energy available for the decay. In stellar environments, the
electron density can be sufficiently high so that the low-energy
states of the electron gas are essentially filled. In such an
environment, there are fewer states available into which a
low-energy electron can be emitted, and therefore $\beta$-decay is
strongly inhibited if maximum kinetic energy available for the decay
is small. For U1F decays, when maximum kinetic energy available for
the decay is small, the inhibition of unique first-forbidden decays
is not as large as the inhibition of allowed, for the same
temperature, density, and $\beta$-decay energy \cite{Bor05}. This
feature is also exhibited in all figures shown.

\section{Conclusions}
In this work the total stellar $\beta$$^{-}$-decay half-lives and
U1F $\beta$-decay rates for a number of neutron-rich nuclei have
been calculated, for the first time, using the pn-QRPA theory. The
temperature and density profiles were relevant to advanced nuclear
burning stages of the presupernova and supernova star. The results
for total $\beta$-decay half-lives are in better agreement with the
experimental results \cite{Fra98,Ame98,Hos05} as compared to the
pn-QRPA calculated half-lives including only allowed GT transitions.
At the same time the calculations suggest that incorporation of rank
0 and rank 1 operators in calculation of first-forbidden transitions
can further improve the results and we plan to work on this in near
future.
\par
The microscopic calculation of U1F $\beta$-decay rates, presented in
this work, could lead to a better understanding of the nuclear
composition and $Y_{e}$ in the core prior to collapse and collapse
phase. The stellar $\beta$-decay rates of waiting point nucleus,
$^{78}$Ni, presented in this work can prove useful for $r$-process
simulation which is a leading candidate for synthesizing heavy
elements in astrophysical sites.

\par
What possible implications can the enhanced U1F $\beta$-decay rates
have for astrophysical scenarios? During the collapse phase allowed
$\beta$-decays become unimportant due to the increased electron
chemical potential which drastically reduces the phase space for the
beta electrons. It is well known that a smaller lepton fraction
disfavors the outward propagation of the post-bounce shock waves, as
more overlying iron core has to be photo-dissociated. It is noted
that the U1F $\beta$-decay rates yield higher Y$_{e}$ in comparison
to (only) allowed $\beta$-decay rates prior to collapse. This might
assist the march of post-bounce shock for a successful supernova
explosion to occur. Further these stronger $\beta$-decay rates can
also assist in a more vigorous URCA process and may lead to cooler
presupernova cores consisting of lesser neutron-rich matter than in
presently assumed  simulations. The reduced  $\beta$-decay
half-lives also bear consequences for site-independent $r$-process
calculations and might result in speeding-up of the $r$-matter flow
relative to calculations based on half-lives calculated from only
allowed GT transitions. However studies on other neutron-rich
isotopic chains including non-unique forbidden contributions, as
well as their relative abundance, are further required to draw more
concrete conclusions and would be taken as a future assignment.
Significant progress is expected, in part, to also come from
next-generation radioactive ion-beam facilities (e.g. FAIR
(Germany), FRIB (USA) and FRIB (Japan)) when we would have access to
measured strength distribution of many more neutron-rich nuclei. The
allowed and U1F $\beta$-decay rates on nickel isotopes were
calculated on a fine temperature-density grid, suitable for
simulation codes, and may be requested as ASCII files from the
authors.

\acknowledgments The authors would like to acknowledge the support
of the Horia Hulubei Foundation through the project
PCE-IDEI-2011-3-0318, Contract no. 58.11.2011.

\onecolumn
\begin{table}[h]
\caption{Comparison of calculated allowed and unique first-forbidden
(U1F) phase space factors for $^{72-75}$Ni for different selected
densities and temperatures in stellar matter. The first column gives
stellar densities ($\rho$Y$_{e}$) having units of g/cm$^{3}$, where
$\rho$ is the baryon density and Y$_{e}$ is the ratio of the
electron number to the baryon number. Temperatures (T$_{9}$) are
given in units of 10$^{9}$ K.}
\begin{tabular}{c|c|cc|cc|cc|cc|}
$\rho Y_{e}$ &T$_{9}$ &\multicolumn{2}{c|}{Phase space ($^{72}$Ni)}
&\multicolumn{2}{c|}{Phase space ($^{73}$Ni)}
&\multicolumn{2}{c|}{Phase space ($^{74}$Ni)}
&\multicolumn{2}{c|}{Phase space ($^{75}$Ni)}
\\
\cline{3-10} && Allowed &
U1F & Allowed & U1F & Allowed &U1F & Allowed & U1F \\

\hline
         & 0.01&    1.7E+04&    6.6E+04 & 9.5E+04 & 4.0E+05 & 8.1E+04 & 4.4E+05 & 4.0E+05 & 1.5E+06 \\
         &    1&    4.2E+08&    1.1E+10 & 2.6E+08 & 1.1E+10 & 1.8E+08 & 7.2E+08 & 5.1E+08 & 1.4E+09 \\
10$^{2}$ &    3&    4.2E+08&    1.1E+10 & 2.6E+08 & 1.1E+10 & 1.8E+08 & 7.2E+08 & 5.1E+08 & 1.4E+09 \\
         &    10&   4.2E+08&    1.1E+10 & 2.6E+08 & 1.1E+10 & 1.8E+08 & 7.2E+08 & 5.0E+08 & 1.4E+09 \\
\hline
         & 0.01&    1.7E+04&    6.6E+04 & 9.3E+04 & 3.9E+05 & 8.0E+04 & 4.4E+05 & 4.0E+05 & 1.5E+06 \\
         &    1&    4.2E+08&    1.1E+10 & 2.6E+08 & 1.1E+10 & 1.8E+08 & 7.2E+08 & 5.1E+08 & 1.4E+09 \\
10$^{6}$ &    3&    4.2E+08&    1.1E+10 & 2.6E+08 & 1.1E+10 & 1.8E+08 & 7.2E+08 & 5.1E+08 & 1.4E+09 \\
         &    10&   4.2E+08&    1.1E+10 & 2.6E+08 & 1.1E+10 & 1.8E+08 & 7.2E+08 & 5.0E+08 & 1.4E+09 \\
\hline
          &0.01&    0.0E+00&    0.0E+00 & 0.0E+00 & 0.0E+00 & 0.0E+00 & 0.0E+00 & 0.0E+00 & 0.0E+00 \\
          &   1&    5.5E+07&    2.5E+09 & 1.3E+07 & 5.3E+09 & 9.2E+06 & 2.0E+08 & 4.2E+07 & 5.6E+08 \\
10$^{10}$ &   3&    5.6E+07&    2.6E+09 & 1.3E+07 & 5.3E+09 & 9.7E+06 & 2.0E+08 & 4.3E+07 & 5.6E+08 \\
          &   10&   7.0E+07&    3.0E+09 & 2.0E+07 & 5.5E+09 & 1.5E+07 & 2.3E+08 & 5.8E+07 & 3.5E+08 \\
\hline
\end{tabular}
\end{table}
\begin{table}[h]
\caption{Same as Table~1 but for $^{76-78}$Ni.}
\begin{tabular}{c|c|cc|cc|cc|}
$\rho Y_{e}$ &T$_{9}$ &\multicolumn{2}{c|}{Phase space ($^{76}$Ni)}
&\multicolumn{2}{c|}{Phase space ($^{77}$Ni)}
&\multicolumn{2}{c|}{Phase space ($^{78}$Ni)}
\\
\cline{3-8} && Allowed &
U1F & Allowed & U1F & Allowed &U1F \\

\hline
         & 0.01&    1.9E+05 & 1.1E+06 & 4.3E+05 & 2.1E+06 & 3.4E+05 & 2.3E+06\\
         &    1&    2.3E+08 & 3.2E+08 & 9.6E+08 & 1.3E+09 & 2.3E+08 & 4.0E+08\\
10$^{2}$ &    3&    2.3E+08 & 3.2E+08 & 9.6E+08 & 1.3E+09 & 2.3E+08 & 4.0E+08\\
         &    10&   2.3E+08 & 3.1E+08 & 9.5E+08 & 1.3E+09 & 2.3E+08 & 4.0E+08\\
\hline
         & 0.01&    1.9E+05 & 1.1E+06 & 4.3E+05 & 2.1E+06 & 3.4E+05 & 2.3E+06\\
         &    1&    2.3E+08 & 3.2E+08 & 9.6E+08 & 1.3E+09 & 2.3E+08 & 4.0E+08\\
10$^{6}$ &    3&    2.3E+08 & 3.2E+08 & 9.6E+08 & 1.3E+09 & 2.3E+08 & 4.0E+08\\
         &    10&   2.3E+08 & 3.1E+08 & 9.5E+08 & 1.3E+09 & 2.3E+08 & 4.0E+08\\
\hline
          &0.01&    0.0E+00 & 0.0E+00 & 0.0E+00 & 0.0E+00 & 0.0E+00 & 0.0E+00\\
          &   1&    2.2E+07 & 6.8E+07 & 1.3E+08 & 5.4E+08 & 2.5E+07 & 5.5E+07\\
10$^{10}$ &   3&    2.3E+07 & 6.9E+07 & 1.4E+08 & 5.5E+08 & 2.5E+07 & 5.7E+07\\
          &   10&   3.0E+07 & 8.1E+07 & 1.7E+08 & 5.7E+08 & 3.3E+07 & 7.1E+07\\
\hline
\end{tabular}
\end{table}


\vspace*{0.01cm}

\vspace*{0.01cm}

\vspace*{0.01cm}

\vspace*{0.01cm}

\vspace*{0.01cm}

\vspace*{0.01cm}

\vspace*{0.01cm}

\end{document}